\documentclass[journal]{IEEEtran}

\usepackage{graphicx}
\usepackage{url}
\usepackage{stfloats}
\usepackage{wasysym}

\begin{document}
\title{Production Performance of the ATLAS Semiconductor Tracker Readout System}
\author{Vasiliki~A.~Mitsou, on behalf of the ATLAS SCT Collaboration%
\thanks{Manuscript received June 18, 2005; revised March 20, 2006.
        This work was supported in part by the EU under the RTN contract:
        HPRN-CT-2002-00292 {\em Probe for New Physics}.}%
\thanks{V. A. Mitsou is with the Instituto de F\'{i}sica Corpuscular (IFIC), CSIC
-- Universitat de Val\`{e}ncia, Edificio Institutos de Investigaci\'{o}n, P.O.
Box 22085, E-46071 Valencia, Spain (phone: +34-963543495; fax: +34-963543488;
e-mail: Vasiliki.Mitsou@cern.ch).}}


\maketitle

\begin{abstract}

The ATLAS Semiconductor Tracker (SCT) together with the pixel and the
transition radiation detectors will form the tracking system of the ATLAS
experiment at LHC. It will consist of 20\,000 single-sided silicon microstrip
sensors assembled back-to-back into modules mounted on four concentric barrels
and two end-cap detectors formed by nine disks each. The SCT module production
and testing has finished while the macro-assembly is well under way. After an
overview of the layout and the operating environment of the SCT, a description
of the readout electronics design and operation requirements will be given. The
quality control procedure and the DAQ software for assuring the electrical
functionality of hybrids and modules will be discussed. The focus will be on
the electrical performance results obtained during the assembly and testing of
the end-cap SCT modules.

\end{abstract}

\begin{keywords}
ATLAS, data acquisition, quality control, silicon radiation detectors.
\end{keywords}

\section{Introduction}

\PARstart{T}{he} ATLAS detector \cite{atlas} is one of the two general-purpose
experiments currently under construction for the Large Hadron Collider (LHC) at
CERN. LHC is a proton-proton collider with a 14-TeV centre-of-mass energy and a
design luminosity of $10^{34}~{\rm cm^{-2}s^{-1}}$. ATLAS consists of the Inner
Detector (ID), the electromagnetic and the hadronic calorimeters, and the muon
spectrometer. The ID \cite{id} is a system designed for tracking, particle
identification and vertex reconstruction, operating in a 2-T superconducting
solenoid. The Semiconductor Tracker (SCT) forms the middle layer of the ID
between the pixel detector and the transition radiation detector.

\begin{figure}[t]
\centering
\includegraphics[width=0.9\linewidth]{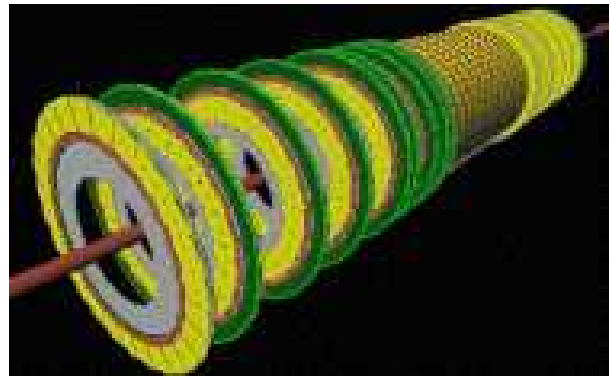}
\caption{Layout of the ATLAS Semiconductor tracker.}
\label{fig:sct}
\end{figure}

The SCT system \cite{id,sct}, depicted in Fig.~\ref{fig:sct}, comprises a
barrel made of four nested cylinders and two end-caps of nine disks each. The
cylinders together carry 2112 detector units ({\em modules}) while 1976 end-cap
modules are mounted on the disks in total. The whole SCT occupies a cylinder of
5.6~m in length and 56~cm in radius with the innermost layer at a radius of
27~cm. It provides a pseudorapidity coverage of up to $\pm2.5$.

The silicon modules \cite{module} consist of one or two pairs of single-sided
p-in-n microstrip sensors glued back-to-back at a 40-mrad stereo angle to
provide two-dimensional track reconstruction. The 285-$\mu{\rm m}$ thick
sensors \cite{sensor} have 768 AC-coupled strips with an $80~\mu{\rm m}$ pitch
for the barrel and a 57--$94~\mu{\rm m}$ pitch for the end-cap modules. Between
the sensor pairs there is a highly thermally conductive baseboard. Barrel
modules follow one common design, while for the forward ones four different
types exist based on their position in the detector.

The readout of the module is based on 12 ABCD3TA ASICs manufactured in the
radiation-hard DMILL process mounted on a copper/kapton hybrid \cite{hybrid}.
The ABCD3TA chip \cite{ABCD} features a 128-channel analog front end consisting
of amplifiers and comparators and a digital readout circuit operating at a
frequency of 40.08~MHz. This ASIC utilizes the binary scheme where the signals
from the silicon detector are amplified, compared to a threshold and only the
result of the comparison enters the input register and the digital pipeline.
The clock and command signals as well as the data are transferred from and to
the off-detector electronics through optical links.

The ID volume will be subject to a fluence of charged and neutral particles
from the collision point and from back-scattered neutrons from the
calorimeters. An estimated fluence at the innermost part of the SCT is
$\sim\!2\times10^{14}$~1-MeV-neutrons$\rm/cm^2$ (or equivalently
$\sim\!3\times10^{14}$~24-GeV-protons$\rm/cm^2$) in ten years of operation. The
SCT has been designed to be able to withstand these fluences \cite{radiation}
and its performance has been extensively studied in beam tests using irradiated
SCT modules \cite{beamtest}.

\section{Electrical requirements}

The LHC operating conditions demand challenging electrical performance
specifications for the SCT modules and the limitations \cite{noise} mainly
concern the accepted noise occupancy level, the tracking efficiency, the timing
and the power consumption. The most important requirements the SCT module needs
to fulfil follow.

\subsubsection{Noise performance}

The total effective noise of the modules results from two principal
contributions; the front-end electronics and the channel-to-channel threshold
matching. The former is the Equivalent Noise Charge (${\rm e^-}$~ENC) for the
front-end system including the silicon strip detector. It is specified to be
less than 1500~${\rm e^-}$~ENC before irradiation and 1800~${\rm e^-}$~ENC
after the full dose of $3\times10^{14}$~24-GeV-equivalent-protons$\rm/cm^2$.
The noise hit rate needs to be significantly less than the real hit occupancy
to ensure that it does not affect the data transmission rate, the pattern
recognition and the track reconstruction. The foreseen limit of
$5\times10^{-4}$ per strip requires the discrimination level in the front-end
electronics to be set to 3.3 times the noise charge. To achieve this condition
at the ATLAS operating threshold of 1~fC, the total equivalent noise charge
should never be greater than 1900~${\rm e^-}$~ENC. Assuming a 3.3-fC median
signal at full depletion that corresponds to a median signal-to-noise ratio of
10:1.

\subsubsection{Tracking efficiency}

In general the tracking performance of a particle detector depends on various
parameters: the radial space available in the cavity, which limits the lever
arm, the strength of the magnetic field, and the intrinsic precision and
efficiency of the detector elements. To this respect a starting requirement is
a low number of dead readout channels, specified to be less than 16 for each
module to assure at least 99\% of working channels. Furthermore no more than
eight consecutive faulty channels are accepted in a module.

\subsubsection{Timing requirements}

For a correct track reconstruction, every hit has to be associated to a
specific bunch crossing. That is translated to a demand for a time-walk of less
than 16~ns, where the time-walk is defined as the maximum time variation in the
crossing of the comparator threshold at 1~fC over a signal range of 1.25 to
10~fC. The fraction of output signals shifted to the wrong beam crossing is
required to be less than 1\%.

\subsubsection{Power consumption}

The nominal values for the power supplies of the ASICs are set as follows:
\begin{itemize}
  \item Analogue power supply: $V_{\rm cc}=3.5~{\rm V}\pm5\%$.
  \item Digital power supply: $V_{\rm dd}=4.0~{\rm V}\pm5\%$.
  \item Detector-bias: high voltage of up to 500~V can be delivered by the ASICs.
\end{itemize}
The nominal power consumption of a fully loaded module is 4.75~W during
operation at 1~fC threshold with nominal occupancy (1\%) and 100~kHz trigger
rate (L1 rate). Including the optical readout, the maximal power dissipation
should be 7.0~W for the hybrid and the heat generated in the detectors just
before thermal run-away should be 2.6~W for outer module wafers and 1.6~W for
inner ones.

\subsubsection{Double pulse resolution}

The double pulse resolution directly affects the efficiency. It is required to
be 50~ns to ensure less than 1\% data loss at the highest design occupancy.

Standard DAQ system and electrical tests, described in the following sections,
aim at verifying the hybrid and detector functionality after the module
assembly and at demonstrating the module performance with respect to the
required electrical specifications.

\section{Data acquisition system}

In all the measurements performed, the ASICs are powered and read out
electrically via the standard SCT DAQ system which contains the following VME
modules:
\begin{itemize}
  \item CLOAC (CLOck And Control): This module generates the clock, fast
  trigger and reset commands for the SCT modules in the absence of the timing,
  trigger and control system.
  \item SLOG (SLOw command Generator): It allows the generation of slow commands
  for the control and configuration of SCT front-end chips for up to six modules.
  It fans out clock and fast commands from an external source (CLOAC).
  Alternatively an internal clock may be selected, allowing SLOG to generate
  clock and commands in stand-alone mode.
  \item MuSTARD (Multichannel Semiconductor Tracker ABCD Readout Device): A unit
  designed to receive, store and decode the data from multiple SCT
  module systems. Up to 12~data streams (six modules) can be read out from one
  MuSTARD.
  \item SCTHV: A prototype high voltage unit providing detector bias to four modules.
  \item SCTLV: A custom-designed low voltage power supply for two silicon modules.
\end{itemize}

\begin{figure}[b]
\centering
\includegraphics[width=\linewidth]{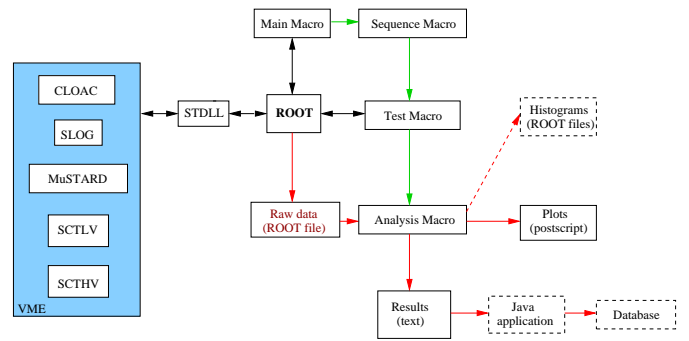}
\caption{Schematic diagram of the SCTDAQ system.}
\label{fig:sctdaq}
\end{figure}

The software package SCTDAQ \cite{sctdaq} has been developed for testing both
the bare hybrids and the modules using the aforementioned VME units. It
consists of a C++ dynamically linked library (STDLL) and a set of ROOT \cite{root}
macros which analyze the raw data obtained in each test and stores the results
in a database \cite{DB}. A schematic diagram of the SCTDAQ is shown in
Fig.~\ref{fig:sctdaq}.

\section{Characterization tests}

\begin{figure*}[!b]
\centering
\includegraphics[angle=-90,width=0.75\linewidth]{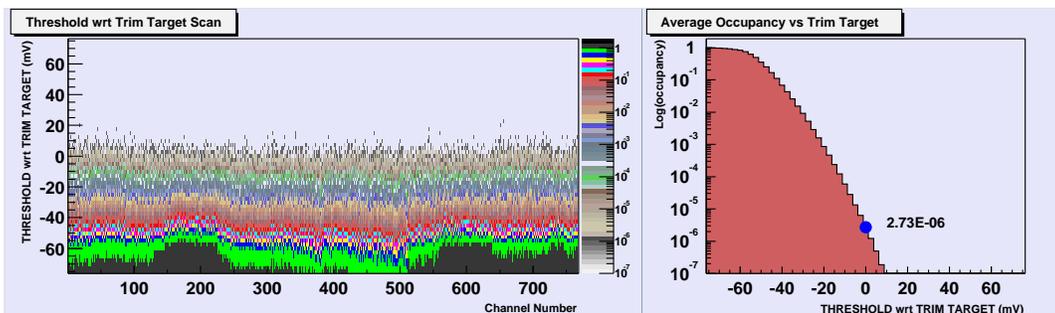}
\caption{Noise occupancy plot for one data stream: occupancy vs.\ channel number
and vs.\ threshold (left); average occupancy  for the stream vs.\ threshold
(right). The threshold is expressed with respect to the 1-fC point (0~mV) as
determined during the trimming procedure.} \label{fig:NoiseOcc}
\end{figure*}

\begin{figure}[t]
\centering
\includegraphics[angle=-90,width=\linewidth,clip=]{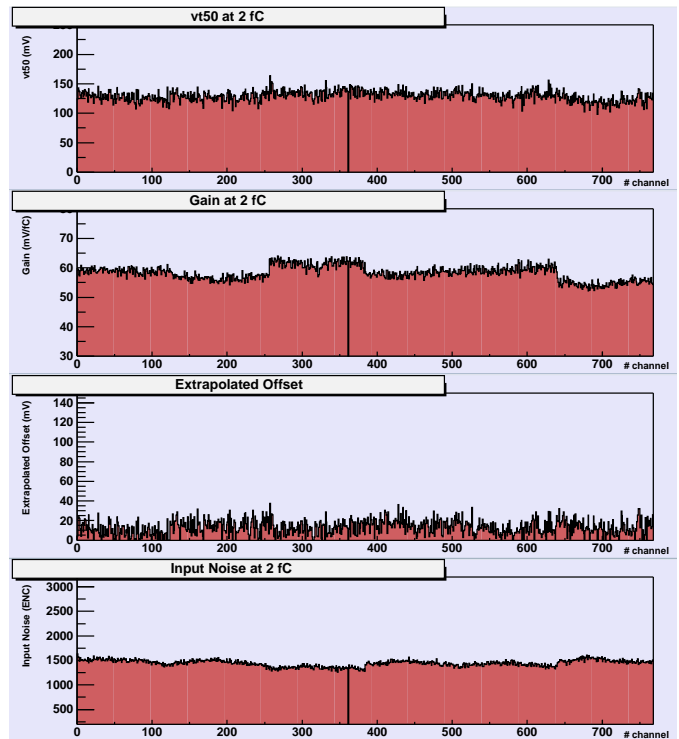}
\caption{Typical set of plots obtained with the Response Curve procedure before
trimming and for one data steam, corresponding to six chips (768 channels).
From top to bottom the $vt_{50}$ value, the gain, the offset and the input
noise are shown for each channel.} \label{fig:3Pgain}
\end{figure}

Every module is characterized to check the functionality and performance
stability and to verify that the specifications are met. Using the internal
calibration circuit to inject charge of adjustable amplitude in the
preamplifier of each channel, the front-end parameters such as gain, noise and
channel-to-channel threshold spread are measured. The characterization sequence
\cite{phillips} includes the following steps:
\begin{itemize}
  \item Digital tests are executed to identify chip or hybrid damage. These
  include tests of the redundancy links, the chip by-pass functionality and
  the 128-cell pipeline circuit.
  \item Optimization of the delay between calibration signal and clock (strobe
  delay) is performed on a chip-to-chip basis.
  \item To minimize the impact of the threshold non-uniformity across the
  channels on the noise occupancy, the ABCD3TA design foresees the possibility
  to adjust the discriminator offset. A threshold correction using a
  digital-to-analog converter (Trim DAC) per channel with four selectable ranges
  (different for each chip) has been implemented in the ASICs. The {\em trimming}
  procedure allows an improved matching of the comparators thresholds; this is
  an important issue for the irradiated modules due to the increase of
  threshold spread with radiation dose.
  \item The gain and electronic noise are obtained channel by channel with
  threshold scans performed for ten different values of injected charge
  ranging from 0.5 to 8~fC (Response Curve procedure; see Fig.~\ref{fig:3Pgain}).
  For each charge injected the corresponding value in mV is extracted as the 50\%
  point ($vt_{50}$) of the threshold scan fitted with a complementary error
  function ($S$-curve). The gain, input noise and offset are deduced from the
  correlation of the voltage output in mV versus the injected charge in fC.
  \item A threshold scan without any charge injection is performed to yield a
  direct measurement of the noise occupancy at 1~fC, as shown in
  Fig.~\ref{fig:NoiseOcc}. The adjusted discriminator offset is applied to ensure a
  uniform measurement across the channels.
  \item A dedicated scan is also executed to determine the time-walk. Setting
  the comparator threshold to 1~fC for each value of injected charge ranging
  from 1.25 to 10~fC a complementary error function is fitted to the falling
  edge of a plot of efficiency versus the setting of the strobe delay to
  determine the 50\%-efficiency point. The time-walk is given by the difference
  between delays calculated for 1.25~fC and for 10~fC injected charge.
\end{itemize}

\begin{figure*}[t]
\centering
\includegraphics[angle=-90,width=0.655\linewidth,clip=]{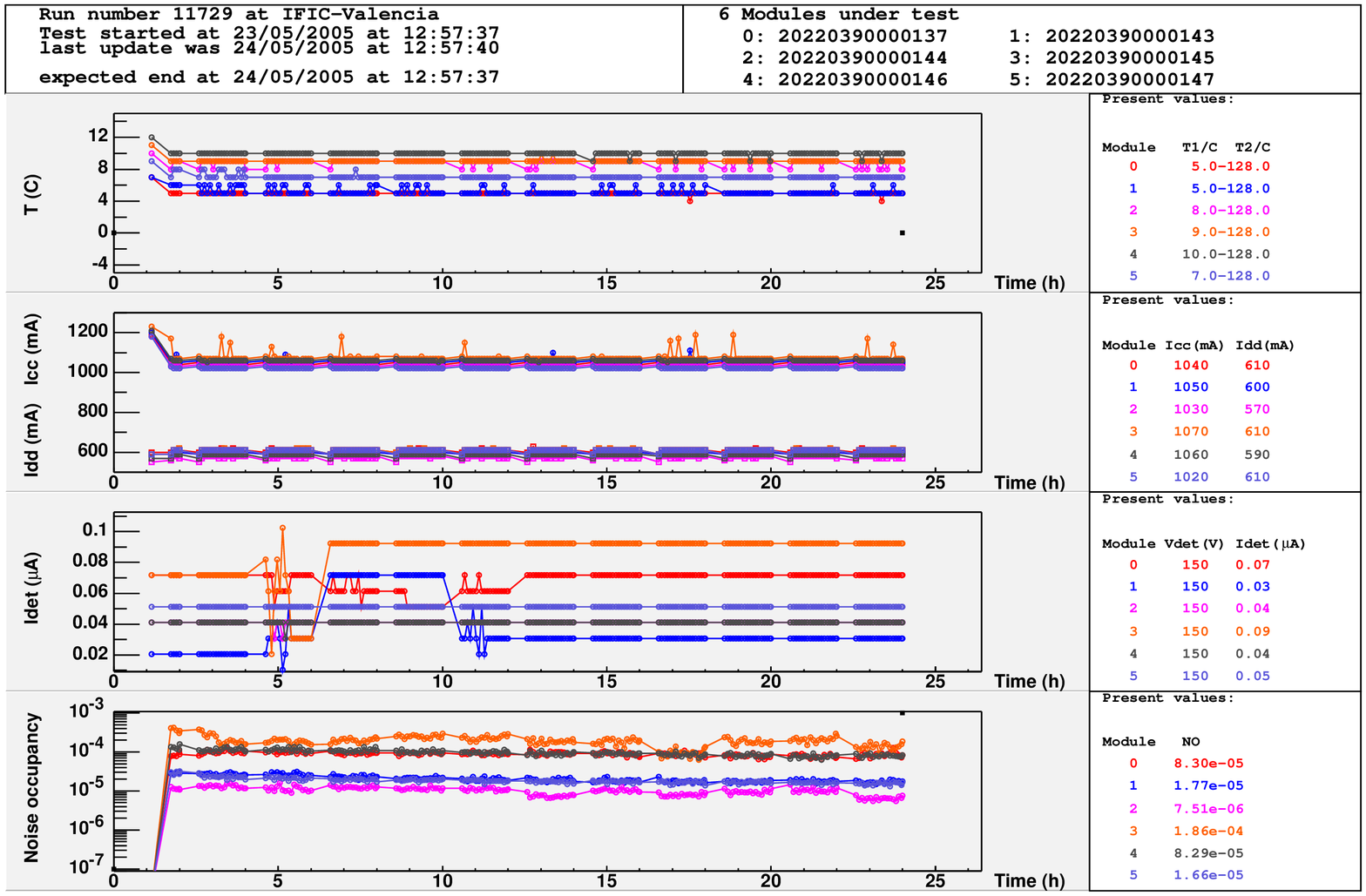}
\caption{Long-term test results for six modules showing from top to bottom:
hybrid temperature; analog ($I_{\rm cc}$) and digital ($I_{\rm dd}$) current;
detector leakage current ($I_{\rm det}$); and noise occupancy as a function of
time.} \label{fig:longterm}
\end{figure*}

As part of the quality assurance test, a long-term test with electrical readout
is also performed. The ASICs are powered, clocked and triggered during at least
18 hours while the module bias voltage is kept at 150~V and its thermistor
temperature is $\rm\sim\!10~^{\circ}C$. The bias voltage, chip currents, hybrid
temperature, the leakage current and the noise occupancy are recorded every
15~min, as shown in Fig.~\ref{fig:longterm}. Moreover, every two hours a test
verifying correct functionality of the module is performed.

A final measurement of the detector leakage current as a function of the bias
voltage ($I-V$ curve) is also performed at $\rm20~^{\circ}C$ to assure that the
current drawn by the whole module is low enough for the safe operation of the
detector. The current values at 150 and 350~V are recorded and compared with
those of previous $I-V$ curve measurements before and after the module
sub-assembly.

During the electrical tests the modules are mounted in a light-tight aluminum
box which supports the modules at the two cooling blocks of the baseboard. The
test box includes a cooling channel connected to a liquid coolant system of
adjustable temperature. The operating temperature is monitored by thermistors
(one for the end-cap and two for the barrel hybrid) mounted on the hybrid. The
box also provides a connector for dry air circulation. Subsequently, the module
test box is placed inside an environmental chamber and it is electrically
connected to the readout system and VME crate. Up to six modules can be tested
simultaneously with this configuration. The grounding and shielding scheme of
the setup is of crucial importance, therefore a careful optimization is
necessary. The tests are carried out at a detector bias of 150~V and at an
operating temperature of $\rm 5-15~^{\circ}C$.

\section{Production modules performance}

All production modules have to pass successfully the aforementioned tests
---long-term test, characterization and leakage current measurement--- as a part
of their quality assurance plan. The hybrids are also tested before assembly
using the same setup and software package. The results presented here
correspond to the end-cap production modules that qualified for assembly onto
disks, which amount to $\sim2000$ (including spares) representing about half of
the total number of SCT modules.

\begin{figure}[t]
\centering
\includegraphics[width=0.88\linewidth]{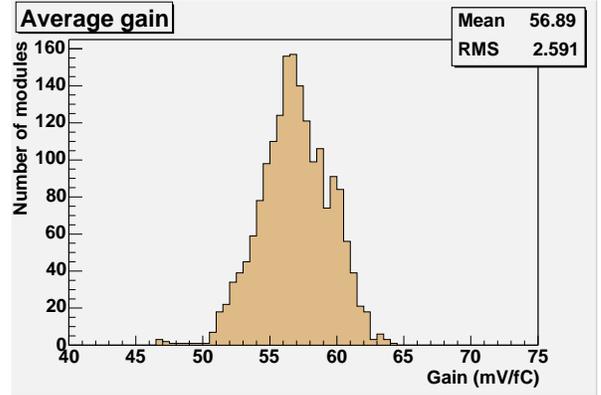}
\caption{Average gain per module for all qualified forward modules.} \label{fig:gain}
\end{figure}

In Fig.~\ref{fig:gain} the average gain per module is shown for all qualified
forward modules. The average gain value is about 57~mV/fC at a discriminator
threshold of 2~fC and it is of the same level as the one obtained from system
tests.

\begin{figure}[t]
\centering
\includegraphics[width=0.88\linewidth]{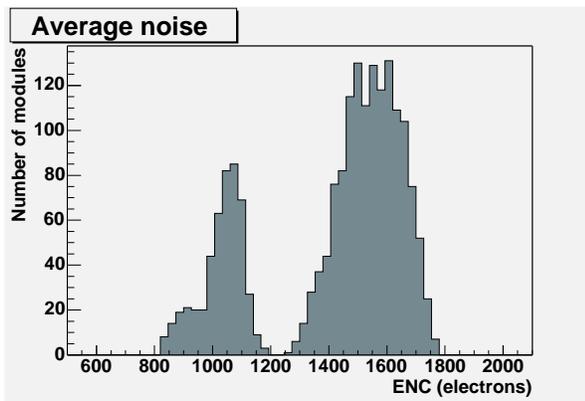}
\caption{Average noise per module for all qualified forward modules.} \label{fig:ENC}
\end{figure}

The noise level per module is shown in Fig.~\ref{fig:ENC}. The two distinct
contributions reflect the difference between {\em short} modules (inner and
short middle) and {\em long} ones (long middle and outer). The former consist
on only one pair of sensors having a strip length of around 6~cm, while the
latter have two detector pairs with a total length of 12~cm, resulting in
higher strip resistance.

\begin{figure}[t]
\centering
\includegraphics[width=0.88\linewidth]{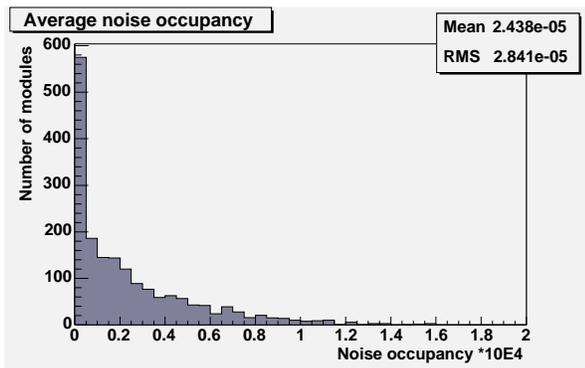}
\caption{Average noise occupancy at 1~fC per module for all qualified forward modules.}
\label{fig:NO}
\end{figure}

An average of 1550 $\rm e^-~ENC$ with an r.m.s.\ of about $100~\rm e^-~ENC$ has
been attained for the long modules. The noise occupancy at a comparator
threshold of 1~fC is measured to be $2.4\times10^{-5}$ on average, i.e.\ twenty
times lower than the requirement of $<5\times10^{-4}$ per strip, as illustrated
in Fig.~\ref{fig:NO}. These values are compatible with the ones acquired from
non-irradiated prototype modules \cite{protonoise}, which also showed that
after irradiation the noise levels although higher do not compromise the
overall detector performance. It should be stressed that the acquired noise
measurements largely depend on the degree of the setup optimization which
generally varies across the testing sites, resulting in a higher than actual
measured value of the module noise. The noise also depends on the temperature
on the hybrid increasing by $\rm\sim\!6~e^-~ENC$ per degree Celsius. Since
under standard conditions at the LHC the modules will operate with a thermistor
temperature near $\rm2~^{\circ}C$, a lower noise level than the one obtained
during quality control tests is expected during running.

\begin{figure}[t]
\centering
\includegraphics[width=0.88\linewidth]{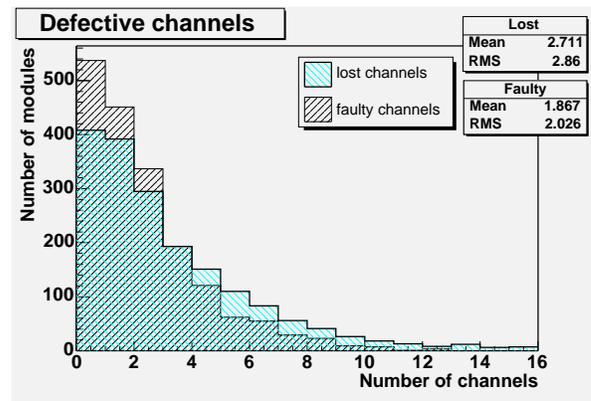}
\caption{Number of lost and faulty channels per module for all qualified forward modules.}
\label{fig:Defects}
\end{figure}

Another aspect of the readout requirements is the number of defective channels
per module. As shown in Fig.~\ref{fig:Defects}, on average less than three
channels per module are {\em lost}, i.e., have to be masked, which represents a
fraction of $1.8\permil$. This category includes dead, stuck, noisy channels,
as well as channels that have not been wire-bonded to the strips and channels
that cannot be trimmed. Other channels exhibit less critical defects such as
low or high gain (or offset) with respect to the chip-average. These {\em
faulty} channels amount to less than two per module ($1.2\permil$). Their
presence is due either to chip defects or defective detector strips (e.g.\
punch-through or short-circuited channels).

As far as the final $I-V$ curves are concerned, the full statistics results
verify the good behavior of the sensors at a high bias voltage. The very few
cases where a problem was observed was either due to detector damage after
assembly or to a defective bias voltage connection on the hybrid. In the latter
case the hybrids were reworked to re-establish the connection.

To recapitulate, only a fraction of about 2.4\% of the tested modules does not
pass at least one electrical characterization test. Most of these modules
exhibit a high number of consecutive faulty channels due to minor damage
(scratch) of module components such as the sensors or the fan-ins.\footnote{The
fan-ins are designed to provide electrical connection between the ABCD chips
and the silicon strips and mechanical support between the hybrid and the
sensors. They are made out of metal tracks deposited on top of a glass
substrate.} The high yield of the electrical tests performed on the production
modules reflects the strict quality control criteria set during the ASICs and
the hybrids selection.

\section{Conclusion}

The results of the systematic electrical tests performed in all SCT forward
production modules demonstrate that they are well within specifications. The
attained gain and the noise performance are compatible with the ones obtained
in several system tests involving detector and electronics prototypes. The
fraction of defective channels per module is kept well below 1\%. The
production of the silicon modules has finished and their mounting onto large
structures (cylinders and disks) is well under way. The whole SCT is expected
to be ready for installation in the ATLAS cavern at the LHC ---together with
the transition radiation detector--- in spring 2006.

\section*{Acknowledgment}

The author would like to thank Carlos Lacasta and Joe Foster for their help in
retrieving the data presented here from the corresponding database and for
useful comments during the preparation of this contribution.

 \IEEEtriggeratref{15}




\begin{thebibliography}{30}

\bibitem{atlas} ATLAS Collaboration, ATLAS Technical Proposal,
  CERN/LHCC/94-43 (1994); \url{http://atlas.web.cern.ch/Atlas/}

\bibitem{id} ATLAS Collaboration, Inner Detector
  Technical Design Report vol.\ I \& II, CERN/LHCC/97-16 \&
  CERN/LHCC/97-17 (1997).

\bibitem{sct} {\em For a recent review see:}
  J.~N.~Jackson  [ATLAS SCT Collaboration],
  Nucl.\ Instrum.\ Meth.\ A {\bf 541} (2005) 89.

\bibitem{module} C.~Lacasta,
  Nucl.\ Instrum.\ Meth.\ A {\bf 512} (2003) 157.

\bibitem{sensor}
  D.~Robinson {\it et al.},
  Nucl.\ Instrum.\ Meth.\ A {\bf 485} (2002) 84.

\bibitem{hybrid} C.~Ketterer,
  IEEE Trans.\ Nucl.\ Sci.\  {\bf 51} (2004) 1134.

\bibitem{ABCD} W.~Dabrowski,
  Nucl.\ Instrum.\ Meth.\ A {\bf 501} (2003) 167.

\bibitem{radiation}
  I.~Mandic  [ATLAS SCT Collaboration],
  IEEE Trans.\ Nucl.\ Sci.\  {\bf 49} (2002) 2888; \\
P.~J.~Dervan  [ATLAS SCT Collaboration],
  Nucl.\ Instrum.\ Meth.\ A {\bf 514} (2003) 163; \\
P.~K.~Teng {\it et al.},
  Nucl.\ Instrum.\ Meth.\ A {\bf 497} (2003) 294; \\
L.~S.~Hou, P.~K.~Teng, M.~L.~Chu, S.~C.~Lee and D.~S.~Su,
  Nucl.\ Instrum.\ Meth.\ A {\bf 539} (2005) 105.

\bibitem{beamtest} Y.~Unno {\it et al.},
  IEEE Trans.\ Nucl.\ Sci.\  {\bf 49} (2002) 1868; \\
F.~Campabadal {\it et al.},
  Nucl.\ Instrum.\ Meth.\ A {\bf 538} (2005) 384.

\bibitem{noise} C.~Lacasta, ``Electrical specifications and expected
performance of the end-cap module,'' ATLAS Project Document, ATL-IS-EN-0008
(2002), \url{https://edms.cern.ch/document/316205/1}

\bibitem{sctdaq} \url{http://sct-testdaq.home.cern.ch/sct-testdaq/sctdaq/sctdaq.html}

\bibitem{root} \url{http://root.cern.ch}

\bibitem{phillips} P.~W.~Phillips and L.~Eklund, ``Electrical tests of SCT hybrids
and modules,'' ATLAS internal document,
\url{http://hepwww.rl.ac.uk/atlas-sct/documents/Electrical_Tests.htm}

\bibitem{DB} C.~Lacasta, F.~Anghinolfi, J.~Kaplon, R.~Szczygiel, W.~Dabrowski, P.~Demierre and D.~Ferrere,
``Production database for the ATLAS-SCT front end ASICs,''
Proc.\ {\it 6th Workshop on Electronic for LHC Experiments, Cracow, Poland,
11-15 Sep 2000} [CERN-2000-010] (2000).

\bibitem{protonoise} C.~Lacasta {\em et al.}, ``Electrical results from
prototype modules,'' ATLAS Project Document, ATL-IS-TR-0001 (2002),
\url{https://edms.cern.ch/document/316209/1}; \\
  P.~W.~Phillips  [ATLAS SCT Collaboration],
  ``System performance of ATLAS SCT detector modules,''
Proc.\ {\it 8th Workshop on Electronics for LHC Experiments, Colmar, France,
9-13 Sep 2002} [CERN-2002-003], p.\ 100--104 (2002).

\end{thebibliography}
\end{document}